\documentclass[lettersize,journal]{IEEEtran}
\usepackage{amsmath,amsfonts}

\usepackage[linesnumbered,ruled,vlined]{algorithm2e}
\usepackage{array}
\usepackage[caption=false,font=normalsize,labelfont=sf,textfont=sf]{subfig}
\usepackage{textcomp}
\usepackage{stfloats}
\usepackage{url}
\usepackage{verbatim}
\usepackage{graphicx}
\usepackage{svg}
\usepackage{cite}
\usepackage{caption}
\usepackage{multirow}
\usepackage{multicol}
\usepackage[utf8]{inputenc}
\usepackage{hyperref}
\begin{document}
	
	\title{VATP360: Viewport Adaptive 360-Degree Video Streaming based on Tile Priority}
	
	\author{Zhiyu Pang,~\IEEEmembership{Nanjing University of Information Science and Technology, 202212490756@nuist.edu.cn}
		\thanks{}
		\thanks{}}
	
	\markboth{Journal of \LaTeX\ Class Files,~Vol.~14, No.~8, August~2021}%
	{Shell \MakeLowercase{\textit{et al.}}: A Sample Article Using IEEEtran.cls for IEEE Journals}

	\maketitle
	
	\begin{abstract}
		360-degree video becomes increasingly popular among users. In the current network bandwidth, serving high resolution 360 degree video to users is quite difficult. Most of the work has been devoted to the prediction of user viewports or tile-based adaptive algorithms. However, it is difficult to predict user viewports more accurately using only information such as user's historical viewports or video saliency maps. In this paper, we propose a viewport adaptive 360-degree video streaming method based on tile priority (VATP360), which tries to balance between the performance and the overhead. The proposed VATP360 consists of three main modules: viewport prediction, tile priority classification and bitrate allocation. In the viewport prediction module, object motion trajectory and predicted user's region-of-interest (ROI) are used to achieve accurate prediction of the user's future viewport.
		Then, the predicted viewport, along with the object motion trajectory, are fed into the proposed tile priority classification algorithm to assign different priorities to tiles, which would reduce the computational complexity of the bitrate allocation module.
		Finally in the bitrate allocation stage, we adaptively assign bitrates to tiles of different priority by reinforcement learning. Experimental results on publicly available datasets have demonstrated the effectiveness of the proposed method. 
	\end{abstract}
	
	\begin{IEEEkeywords}
		360-degree video, viewport prediction, tile priority classification, reinforcement learning
	\end{IEEEkeywords}
	
	\section{Introduction}
	\IEEEPARstart{R}{ecent} years, the rapid development of VR devices and advancements in video capture technology have led to the production of a significant number of immersive video streams, including 360-degree video. Notably, popular video platforms like YouTube, Facebook, ARTE, and Vimeo are actively promoting 360-degree video services\cite{r1,r2}. The 360-degree video mainly revolves around gaming and entertainment, as well as other domains such as education, immersive projection venues, infotainment, and sports\cite{r2,r3,r4}. 
	
	Different from traditional video streaming, 360-degree video is spatially divided into smaller video fragments known as tiles\cite{r2} before transmission. This division occurs after the projection of the video from a 3D sphere to a 2D plane, such as the equirectangular projection\cite{r5,r6}. 
	Then, the tile-based adaptive streaming schemes and viewport prediction schemes are used to assign different bitrates to tiles within the available network bandwidth. 
	Early researcher\cite{r8} adopted a strategy of dividing the entire 360-degree video into multiple video chunks and transmitting them using an equivalent quality (bit rate) similar to conventional video streaming\cite{r9}. However, this approach is suboptimal, since 360-degree videos typically require much higher bit rates than conventional videos. Later, attempts to allocate higher bit rates to foreground views and lower bit rates to background views are proposed\cite{r10}, which is still ineffective in utilizing the network bandwidth. Because the approach is non-adaptive and cannot dynamically adapt to the real network environment. More recently, researchers\cite{r11,r12,r13,r14,r15,rr1} have explored the use of Deep Reinforcement Learning (DRL) to enable adaptive transmission of tiles, leading to effective and efficient utilization of available network resources.
	Besides, users only view a small portion of the 360-degree scene (viewport) in actual. The most network bandwidth is taken up by the unviewed content\cite{r3}. Therefore, it is important to anticipate the user's future viewport. Some researchers\cite{r16,r17,r18,r19} have used the user's historical viewport and the saliency map of the video to predict the user's future viewport by linear methods or long and short-term memory networks (LSTM). In addition, some other researchers have proposed tiling schemes with different rules\cite{r31,rr3} and sizes\cite{rr4,rr5} according to different video contents, as well as taking into account the preferences of different users\cite{r20,r21} and simplifying the complexity of models\cite{r22,r23,r24}.
	
	However, it is acknowledged that achieving highly accurate viewport prediction and adaptive streaming of 360-degree video in realistic scenarios poses significant challenges\cite{r3}. At present, most of the studies are deficient and face multiple challenges. Firstly, existing viewport prediction methods have difficulty in achieving high accuracy when dealing with different types of videos because they do not take into account more information that may affect the user's viewing behaviour. Secondly, although viewport prediction errors are unavoidable, and some of them even errors can cause serious degradation of user QoE, the errors can be remedied by certain methods. Most of the existing work does not have an effective method to remedy the negative impact caused by viewport prediction errors. Thirdly, the current use of reinforcement learning to solve the problem of 360-degree video streaming can achieve great results, but it possesses high complexity and consumes lots of network resources. The need for more flexible solutions to these challenges is therefore the motivation for our work.

	Given the issues identified in the previous works, we propose a novel tile priority-based viewport adaptive streaming scheme for 360-degree videos. Our model comprises three key modules: the viewport prediction module, the tile priority classification module, and the bitrate allocation module, as illustrated in Fig. 1. The proposed viewport prediction module integrates the saliency of video content, the trajectory of the object, and the user’s historical viewport to predict user's future viewport. Additionally, the complexity of the model is significantly reduced through the utilization of the proposed tile priority classification algorithm. This algorithm effectively manages the trade-off between performance and overhead in the 360-degree adaptive streaming solution. Then, A3C is used in the bitrate allocation module to assign different bitrates to tiles of different priority, given the network status. 
	In summary, the main contributions of this paper are as follows:
	
	\begin{itemize}
		\item We use various factors affecting the user's viewing behaviour, such as video saliency map, user's head movement trajectory and video object tracking trajectory, to predict the user's future viewport, which greatly improves the prediction accuracy.
		\item Our proposed tile priority classification algorithm takes the predicted viewport and the video object tracking trajectory as inputs, and assigns priorities to the tiles in different regions of the video frame by combining the specific content of the video and the user's viewing behaviours, which can effectively deal with the negative impact caused by the viewport prediction error.
		\item In the bitrate allocation stage, we consider the number of each tile priority as an input to the A3C network, which optimises the decision space for reinforcement learning and reduces the overhead of network resources.
		\item We extensively evaluate our proposed model over publicly available datasets. The results of experiments demonstrate that our algorithm outperforms other benchmark algorithms significantly.
	\end{itemize}
	
	The rest of this paper is structured as follows. In Section \uppercase\expandafter{\romannumeral 2}, we summarize related works in 360-degree video streaming. In Section \uppercase\expandafter{\romannumeral 3}, we introduce the overall model in detail, as well as the design of each module. In Section \uppercase\expandafter{\romannumeral 4}, we present the dataset and the associated setup for the experiments. In Section \uppercase\expandafter{\romannumeral 5}, we conduct extensive evaluation experiments of the proposed model on publicly available datasets and analyse the results of the experiments. Finally, in Section \uppercase\expandafter{\romannumeral 6} and \uppercase\expandafter{\romannumeral 7}, we conclude the whole paper and future work.
	
	\section{Related work}
	Currently, research solutions on 360-degree adaptive streaming are mainly classified into tile-based streaming, viewport prediction and tile partition. In this section, we will focus on some related work in these three classes of solutions.
	
	\textbf{Tile-based streaming:} In the tile-based streaming solutions, the selection and transmission of tiles are predominantly determined by the user's viewport. The end device receives only the specific region of the video frame that corresponds to a field of view equal to or larger than the viewport range\cite{r11,r12}. In the majority of these studies\cite{r11,r13,r14}, high bitrates are adaptively assigned to tiles within the viewport, while tiles outside the viewport receive low or no bitrates. A few studies\cite{r12} incorporate salient and dynamic video maps, along with the user's historical viewport traces, as input to the network. In\cite{r15}, The authors propose that 360HRL divides video transmission and tile selection into two stages of training and allows for reselection of erroneous tiles to mitigate the effects of viewport prediction errors. Jiang et al.\cite{rr1} propose a tile prefetching strategy that dynamically updates the tiles based on the latest prediction results to further reduce the adverse effects of prediction errors.
	
	\textbf{Viewport prediction:} For the viewport prediction, previous research has utilized the user's historical viewport data\cite{r17,r18,r19} and the video's saliency map\cite{r16} to predict the user's future viewport using either linear approaches or LSTMs. In\cite{r25}, a graph convolutional network (GCN) based on the approach presented in\cite{r26} is employed to predict saliency maps, with high-scoring regions serving as predicted viewports. Another approach, as discussed in\cite{r23}, involves viewport prediction using clustering methods based on viewer viewport patterns. Yaqoob et al.\cite{rr2} designed two prediction mechanisms to process viewports in both vertical and horizontal directions, then combined the results to interactively select tile. Additionally, researchers in the field of computer vision have explored techniques such as object tracking to predict user viewports\cite{r27}. 
	
	\textbf{Tile partitioning:} In 360-degree video streaming, the video chunk can be further divided into tiles of equal/unequal shape and size to accurately adjust the bitrate of different regions of the video frame and reduce coding redundancy. The work\cite{r31} calculates the distance from the centre of each tile to the viewport position and gradually increases the bitrate of the different position tiles in a decreasing order. In\cite{rr3}, the authors proposed a heuristic algorithm that dynamically selects the coverage of the viewport and its edge regions and assigns different weights to the tiles in these regions. Distinct from\cite{r31} and\cite{rr3}, \cite{rr4} proposes a variable-size tile partitioning and coding scheme based on the specific content of the video. In order to trade-off the viewport prediction error and the waste of network resources, some researchers\cite{rr5} have proposed a method of tiling with different variable number of tiles, e.g., 4 × 4, 6 × 6, and 6 × 8 divisions.
	
	Our approach is innovative in viewport prediction and tile partitioning. Our model takes into account multiple factors affecting user viewing behaviour for viewport prediction, while the negative impact of viewport prediction errors is reduced by a tile prioritization classification algorithm, and the complexity of the model is effectively reduced.
	
	
	\section{System Overview}
	This Section presents a comprehensive description of each module in the proposed model. Our model comprises three main modules as illustrated in Fig. 1: the viewport prediction module, the tile priority classification module, and the bitrate allocation module. The viewport prediction module uses spatio-temporal information from the video saliency map, the user's head movement trajectory and the orientation information of objects to predict the user's future viewport. The tile classification module integrates the predicted viewports with the results of object detection to assign the priority level to each tile. Subsequently, the bitrate allocation module takes the outputs of the tile classification module in conjunction with network status and video features as inputs. And the reinforcement learning A3C architecture is employed to optimize the allocation of bitrates to individual tiles, with the objective of maximizing the QoE.
	
	\begin{figure*}[!t]
		\centering
		\includegraphics[width=\textwidth]{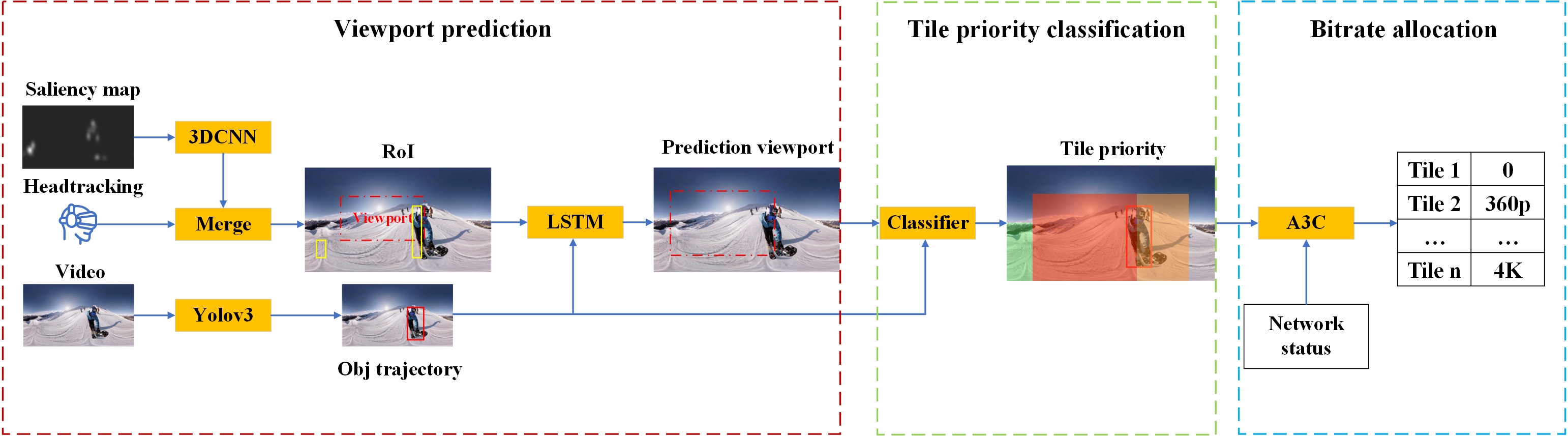}
		\captionsetup{format=plain, justification=justified}
		\caption{The framework of proposed Viewport adaptive 360-degree video streaming based on tile priority.}
		\label{fig_1}
	\end{figure*}
	
	In our work, we choose the equirectangular projection method (ERP)\cite{r7} for preprocessing 360-degree videos. We split the 360-degree video temporally into multiple video chunks of the same duration $C=\left\{c_{0},c_{1},...,c_{m}\right\}$. For each video chunk we further split it spatially into multiple tiles of the same size $T_{m}=\left\{t_{m,0},t_{m,1},...,t_{m,n}\right\}$ for bitrate allocation.
	
	\subsection{Viewport prediction}
	Prediction of user's viewport is important for 360-degree video streaming, where mis-prediction will lead to reduction in the user's QoE\cite{r12,r28}. As shown in Fig. 1 and Algorithm 1, we first perform feature extraction on the video saliency map by 3DCNN, and then integrate it with the user's head movement trajectory as the user's possible region of interest (RoI). Finally, the user's head movement in the future period is predicted by an LSTM network. For the m-th video block, the user's viewport at the i-th moment can be obtained from Eq. (1),
	\begin{equation}
		V_{i}=LSTM(S_{i-1},U;\phi)
	\end{equation}
	where $S_{i-1}=\left\{s_{0},s_{1},...,s_{i-1}\right\}$ is the set of user head movement trajectories from $0$ to $i-1$, $U_{i}=\left\{u_{1},u_{2},...,u_{i}\right\}$ is the set of video saliency maps from $1$ to $i$, and $\phi$ is the parameter of the LSTM network.
	
	\begin{algorithm}
		\SetAlgoLined
		\LinesNotNumbered
		\caption{Viewport Prediction Refinement Algorithm}
		\label{alg:viewport_prediction}
		\SetKwInOut{Input}{Input}
		\SetKwInOut{Output}{Output}
		
		\Input{$V$, $\phi$, $T_m=\{t_{m,0},t_{m,1},...,t_{m,n}\}$}
		\Output{$T_{v}$, $T_{obj}$}
		
		$V_{i}=LSTM(S_{i-1},U;\phi)$
		
		$T_{\text{obj}}=\{t_{m,0},t_{m,1},...,t_{m,j}\}$ 
		
		$T_{v}=\{t_{m,0},t_{m,1},...,t_{m,k}\}$ 
		
		\If{$T_{obj} \cap T_{v} \neq \emptyset$}{
			\For{$i \gets 0$ to $j$}{
				add $t_{m,i}$ to $T_{v}$
			}
		}
		
		\Return{$T_{v}$, $T_{obj}$}
	\end{algorithm}
	In order to improve the accuracy of the predictions and to reduce the detrimental effects of prediction errors, we also consider object trajectories in the video, as users are more likely to be attracted to relevant objects in the video content. We utilize the YOLOv3\cite{r29} model to detect relevant objects in the video. As shown in Fig. 2, we define two sets of tiles, $T_{obj}$ and $T_{v}$, to hold the tiles that lie within the object box and the predicted viewport respectively. For the tiles that lie within the object box and at the edge of the predicted viewport, we add them to $T_{v}$ to adjust the range of the predicted viewport.
	
	\begin{figure}[!t]
		\centering
		\includegraphics[width=\columnwidth]{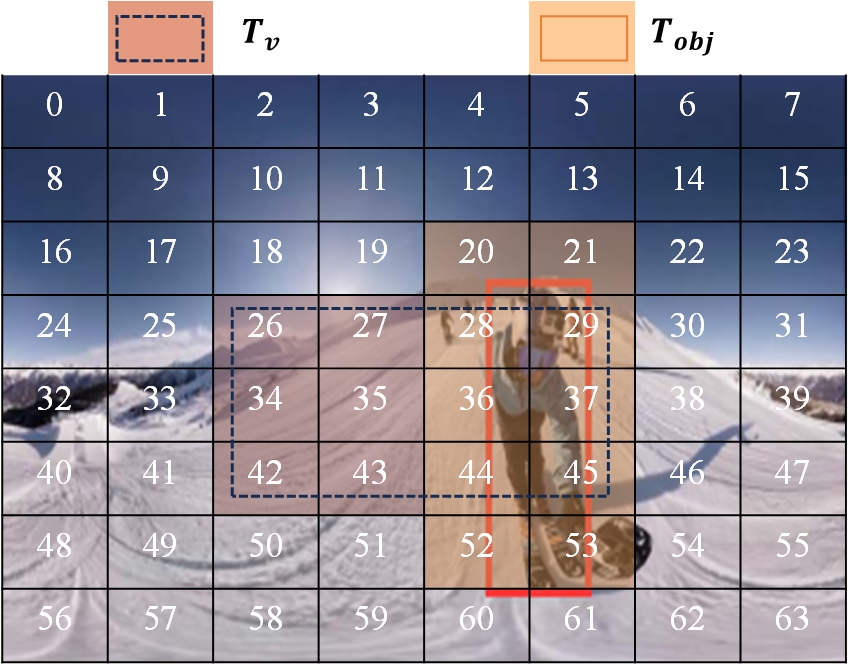}
		\captionsetup{format=plain, justification=justified} 
		\caption{We merge the set of tiles contained in the viewport and the object box that have overlapping parts with the viewport as the final predicted viewport.}
		\label{fig:viewport}
	\end{figure}

	\subsection{Tile classification}
	In order to compensate for the negative impact of prediction error and to reduce the complexity of the model, we propose a classification approach to categorize the tiles based on predicted viewports and object trajectories, assigning them different priorities. While some existing studies \cite{r30,r31} also consider assigning different bitrates to tiles based on their locations, they do not classify the tiles in conjunction with specific video content. We will demonstrate the effectiveness of our proposed tile prioritisation algorithm in the ablation experiments in Section 4.
	
	\begin{figure}[!t]
		\centering
		\includegraphics[width=\columnwidth]{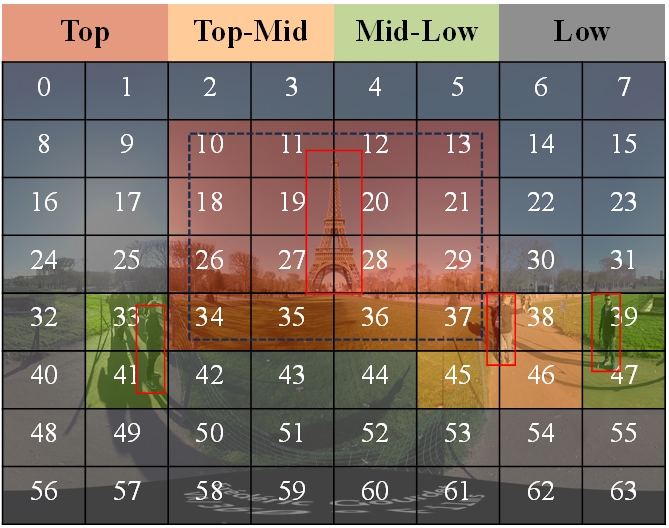}
		\captionsetup{format=plain, justification=justified} 
		\caption{Tile priority classification. We assign different priorities to different areas of the tile according to the predicted viewport (black dashed box) and the object detection box (red rectangle box). The priority of each tile is denoted by its background color.}
		\label{fig:tile}
	\end{figure}
	
	\begin{algorithm}
		\LinesNotNumbered
		\SetAlgoLined
		\caption{Tile Priority Classification Algorithm}
		\label{alg:tile_priority}
		\SetKwInOut{Input}{Input}
		\SetKwInOut{Output}{Output}
		
		\Input{$W_m=\{0,0,0,0\}$, $T_{obj}$, $T_{v}$}
		\Output{$W_{m}$}
		\For{$i \gets 0$ to $m$}{
			\If{$t_{m,i} \in T_v$}{
				$w_{\text{top}}=w_{\text{top}}+1$
			}
			\ElseIf{$t_{m,i} \in T_{\text{obj}}$ and $T_{\text{obj}} \cap T_v \neq \emptyset$}{
				$w_{\text{top-mid}}=w_{\text{top-mid}}+1$
			}
			\ElseIf{$t_{m,i} \in T_{\text{obj}}$ and $T_{\text{obj}} \cap T_v = \emptyset$}{
				$w_{\text{mid-low}}=w_{\text{mid-low}}+1$
			}
			\Else{
				$w_{\text{low}}=w_{\text{low}}+1$
			}
		}

		\Return{$W_{m}$}
	\end{algorithm}
	
	We define four tile priorities: highest (Top), second highest (Top-Mid), second lowest (Mid-low), and lowest (Low). For each video block, we construct a weight matrix $W_{m}=\left\{w_{top},w_{top-mid},w_{mid-low},w_{low}\right\}$ that represents the number of tiles for each priority level. As shown in Algorithm 2 and Fig. 3, we assign the highest priority to tiles within the predicted viewport, the next highest priority to tiles around the predicted viewport and within the object box, the next lowest priority to tiles outside the predicted viewport but within the object box, and the lowest priority to the remaining tiles.

	\subsection{Bitrate allocation}
	We deployed the A3C model to dynamically allocate bitrates for tiles of 360-degree video. It is worth mentioning that instead of assigning bitrates to individual tiles, we assign them according to the priority levels. We formulate adaptive bit rate selection as a Markov Decision Process. The process contains state, action and reward. We will specify the design of each part as follow.
	
	\textbf{State:} The current network status and information about the video are considered. For the state at time t:
	
	$\left[h_{t-1},h_{t-2},...,h_{t-x}\right]$: Network throughput when downloading several previous tiles, where $x$ indicates the number of tiles that have been downloaded.
	
	$\left[\sigma_{t-1},\sigma_{t-2},...,\sigma_{t-x}\right]$: Time taken to download the previous $x$ tiles.
	
	$\left[p_{0},p_{1},...,p_{n}\right]$: The probability that each tile will be viewed by a user.
	
	$\left[q_{0},q_{1},...,q_{N}\right]$: The download code rate level that can be selected for each priority tile. $q=0$ means that this type of tile will not be downloaded.
	
	$\left[l_{top},l_{top-mid},l_{mid-low},l_{low}\right]$: The download bitrate level for each previous priority tile.
	
	$\left[w_{top},w_{top-mid},w_{mid-low},w_{low}\right]$: Number of tiles per priority level.
	
	$\left[\alpha_{top},\alpha_{top-mid},\alpha_{mid-low},\alpha_{low}\right]$: The number of tiles remaining to be downloaded for each priority level.
	
	$b_{t}$: Playback buffer occupancy.
	
	The network structure are shown in Fig. 4. We used five 1DCNN units for $h_{t}$, $\sigma_{t}$, $p_{n}$, $q_{N}$, $l_{t}$ and three linear units for $w_{i}$, $\alpha_{i}$, $b_{t}$ for feature extraction respectively. And the other steps are similar to those in the related work\cite{r9,r12}.
	
	\textbf{Action:} To simplify the action space of the RL agent, we employ a tile priority classification algorithm, which assigns bitrates to tiles based on four priority levels. In our study, the video is encoded at six different bit rates, offering six potential choices. When the RL agent selects an action from this set of actions, it signifies a switch to the corresponding bitrate level for the next tile in the streaming process.
	
	\begin{figure}[!t]
		\centering
		\includegraphics[width=\columnwidth]{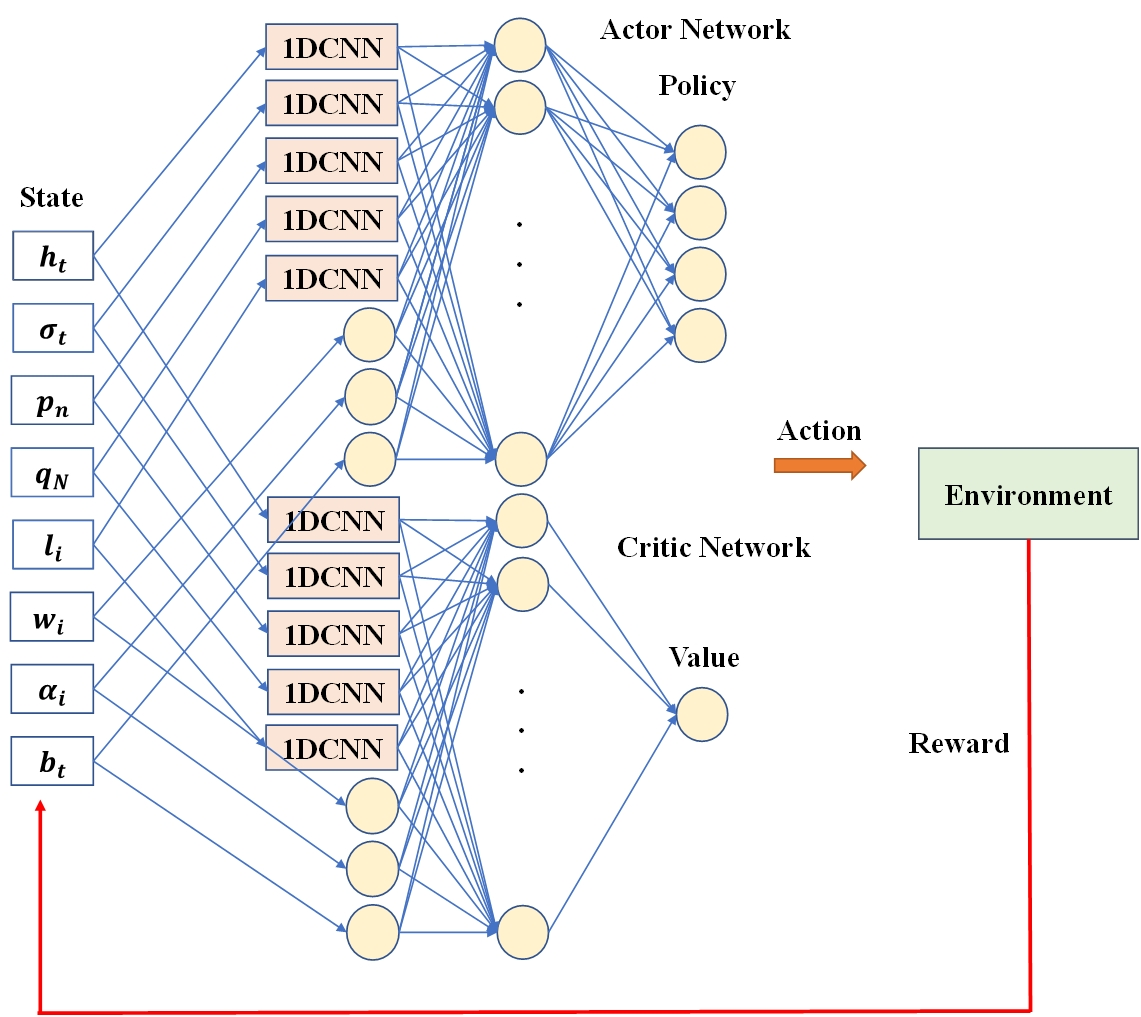}
		\captionsetup{format=plain, justification=justified}
		\caption{The Network structure for bitrate allocation.}
		\label{fig_2}
	\end{figure}
	
	\textbf{Reward:} In our study, we define the RL reward function based on four user QoE metrics:
	
	\textit{Average Viewport Quality:} For the video chunk $c_{m}$, the average quality of all tiles within the viewport during its playback time is calculated as follows.
	\begin{equation}
		QoE_{1}=\frac{1}{n}\sum_{i=1}^{f_{m}}\left({\frac{\sum_{j=1}^{n}{B_{m,j}O_{i,j}}}{\sum_{j=1}^{n}{B_{m,j}}}}\right)
	\end{equation}
	where $f_{m}$ is the number of frames of the video chunk $c_{m}$. $B_{m,j}$ is the playback quality of the j-th tile in the m-th chunk, $O_{i,j}$ indicates whether the tile of the current frame is inside the predicted viewport, $O_{i,j}=1$ indicates that the tile is inside the predicted viewport, otherwise $O_{i,j}=0$.
	
	\textit{Rebuffering Time:} For the video chunk $c_{m}$, we position the rebuffering time as the sum of the rebuffering times for each tile in that video chunk.
	\begin{equation}
		QoE_{2}=\sum_{j=1}^{n}{d_{m,j}}
	\end{equation}
	Where $d_{m,j}$ is the duration of the j-th tile re-buffer when downloading the video chunk $c_{m}$.
	
	\textit{Interframe Quality Smoothness:} For the video chunk $c_{m}$, we hoped to minimise the variation in quality of the viewport between frames to provide a better viewing experience for the user.
	\begin{equation}
		QoE_{3}=\frac{1}{n}stdDev\left(\frac{\sum_{j=1}^{n}{B_{m,j}O_{i,j}}}{\sum_{j=1}^{n}{B_{m,j}}},O_{i,j}=1\right)
	\end{equation}
	
	\textit{Quality Smoothness Across Chunk:} We consider the smoothness of the quality of the viewport between the video chunks $c_{m}$ and $c_{m-1}$ as one of the factors influencing the user QoE.
	\begin{equation}
		\begin{split}
			QoE_{4}&=\frac{1}{n}\mid\sum_{i=1}^{f_{m}}{\left({\frac{\sum_{j=1}^{n}{B_{m,j}O_{i,j}}}{\sum_{j=1}^{n}{B_{m,j}}}}\right)}-\\
			&\sum_{i=1}^{f_{m-1}}{\left({\frac{\sum_{j=1}^{n}{B_{m-1,j}O_{i,j}}}{\sum_{j=1}^{n}{B_{m-1,j}}}}\right)}\mid
		\end{split}
	\end{equation}
	
	In summary, for the video chunk $c_{m}$, the overall QoE can be modelled as:
	\begin{equation}
		QoE=\mu_{1}QoE_{1}-\mu_{2}QoE_{2}-\mu_{3}QoE_{3}-\mu_{4}QoE_{4}
	\end{equation}
	where $\mu_{1}$, $\mu_{2}$, $\mu_{3}$, $\mu_{4}$ are the weights of each QoE component.
	
	\textbf{Training Method:} We train through the asynchronous dominance of reinforcement learning A3C algorithm. The Actor network updates the parameters in the network by means of a policy gradient, and the Critic network updates itself and the Actor network by means of a value function guide. Specifically, the Actor network (policy) receives state from the environment and selects the action to be performed $\pi_{\theta}\left(s,a\right)$. Meanwhile, the Critic network (value function) receives the state and rewards generated by previous interactions and uses this information to compute TD (Temporal difference) errors to guide the training process. This approach weighs the variance of reducing the strategy gradient against the bias of introducing a value function approach.
	
	In contrast to traditional AC algorithms, A3C combines dominance updates with Actor-Critical algorithms and relies on a network of asynchronous update strategies and value functions trained in parallel on multiple processing threads. For the strategy $\pi_{\theta}\left(s_{t},a_{t}\right)$ , A3C calculates its difference from the average action by means of a dominance function. The dominance function is defined as:
	\begin{equation}
		A^{\pi_{\theta}}\left(s_{t},a_{t}\right)=Q^{\pi_{\theta}}\left(s_{t},a_{t}\right)-V^{\pi_{\theta}}\left(s_{t}\right)
	\end{equation}
	where $V^{\pi_{\theta}}\left(s_{t}\right)$ is the estimated value of the value function at the current state  after the Actor network has made an action based on the policy $\pi_{\theta}\left(s_{t},a_{t}\right)$. $Q^{\pi_{\theta}}\left(s_{t},a_{t}\right)$ is the value at the current state  after making a particular action $a_{t}$. For the strategy $\pi_{\theta}\left(s_{t},a_{t}\right)$, the value function can be calculated as the expected discounted return and also as the output of the Critic network.
	\begin{equation}
		V^{\pi_{\theta}}\left(s_{t}\right)=E_{\pi\left(s_{t}\right)}\left[r_{t}+\gamma V^{\pi_{\theta}}\left(s_{t+1}\right)\right]
	\end{equation}
	Where $r_{t}$ is the value of the action $a_{t}$ and $\gamma$ is the discount factor.
	
	Thus, according to the dominance function, for the parameter $\theta_{a}$ , the Actor network can be updated in the following way.
	\begin{equation}
		\begin{split}
			\theta_{a}\leftarrow\theta_{a}&+\alpha_{a}\sum_{t}{\nabla_{\theta}\log{\pi_{\theta}\left(s_{t},a_{t}\right)}}A\left(s_{t},a_{t}\right)+\\
			&\beta\nabla_{\theta}H\left(\pi_{\theta}\left(\cdot\mid s_{t}\right)\right)
		\end{split}
	\end{equation}
	$\alpha_{a}$ is the learning rate of the Actor network, and $\nabla_{\theta}\log{\pi_{\theta}\left(s_{t},a_{t}\right)}$ indicates the direction in which the network parameters are updated according to the strategy gradient. To encourage the exploration of the intelligence to learn a better strategy, a cross-regularisation term $\nabla_{\theta}H\left(\pi_{\theta}\left(\cdot\mid s_{t}\right)\right)$ is added to aid training. $\beta$ is a dynamic parameter that is set to a large value at the beginning of training and then decreases over time.
	
	For the Critic network, the parameter $\theta_{c}$ is updated as follows:
	\begin{equation}
		\theta_{c}\leftarrow\theta_{c}+\alpha_{c}\sum_{t}{\nabla_{\theta}\left(r_{t}+\gamma V^{\pi_{\theta}}\left(s_{t+1};\theta_{v}\right)-V^{\pi_{\theta}}\left(s_{t};\theta_{v}\right)\right)^2}
	\end{equation}
	$\alpha_{c}$ is the learning rate of the Critic network and $V^{\pi_{\theta}}\left(s_{t};\theta_{v}\right)$ is the output of the Critic network. This allows us to calculate an estimate of the dominance function $A\left(s_{t},a_{t}\right)$. 
	\begin{equation}
		A^{\pi_{\theta}}\left(s_{t},a_{t}\right)\approx r_{t}+\gamma V^{\pi_{\theta}}\left(s_{t+1};\theta_{v}\right)-V^{\pi_{\theta}}\left(s_{t};\theta_{v}\right)
	\end{equation}
	
	It is crucial to note that we are only making bitrate decisions through the Actor network, and the Critic network is only involved in helping to train the Actor network and will not be a direct part of the bitrate decision process. So, when deploying the algorithm in the real world, we do not use the output of the Critic network.
	
	\section{Experimental Settings}
	\textbf{Dataset: }For the viewport prediction module, we utilized two publicly available datasets comprising 360-degree videos and user head tracking logs\cite{r32,r33}, and a salient graph dataset\cite{r34} containing 24 360-degree videos. These datasets have been widely used in various prominent studies\cite{r21,r27,r28}. In our experiments, we selected the initial 60 seconds of data from all videos.
	
	Regarding the bitrate allocation module, we employed 50 network bandwidth traces from the publicly available HSDPA dataset\cite{r35} that represents 3G networks as our experimental data. To simulate the prevalent 4G network, we extended the data from the 3G network to cover the traffic range typically observed in 4G networks, maintaining the same proportional relationship.
	
	\textbf{Data Pre-processing: }In the case of 360-degree videos, the initial 60 seconds of each video were chosen for frame-by-frame projection and subsequent tiling. To process the user head tracking logs, we transformed the unit quaternion into a rectangular viewport that aligns with the user's viewport, employing the method described in\cite{r27}. We divided both the videos and user viewports into three distinct groups, with each group representing a different video type and a distinct preferred viewership demographic.
	
	
	
	
	
	
	
	
	
	\textbf{Hyper Parameter:}
	We set the size of each video block to 1 second and divide each frame into 8 × 8 tiles. In the viewport prediction module, we set the number of cells in the LSTM to 8 and the number of layers to 2. In the bitrate allocation module, we provide 6 bitrate download levels, including 0 Mbps (no transmission), 1 Mbps (360p), 5 Mbps (720p), 8 Mbps (1080p), 16 Mbps (2K) and 35 Mbps (4K). For the RL, we set the discount factor $\gamma=0.99$, the learning rate $\theta_{a}=0.0001$, $\theta_{c}=0.001$ and the crossover factor $\beta$ decreases gradually from 1 to 0.1. For the QoE weights, we set three different combinations of weights: (1, 1, 1, 1), (1, 4, 1, 1), (1, 1, 4, 4). The default QoE weight is set to (1, 1, 1, 1).
	
	\section{Experiment Evaluation}
	In this Section, we provide a comprehensive analysis of all the experimental results presented in this paper. Initially, we compare the accuracy of the viewport prediction achieved by our proposed method against various baseline methods. Subsequently, we assess the QoE for each 360-degree video under three different network bandwidth scenarios. To validate the effectiveness of our approach, we compare it with other baseline approaches regarding each component of the average QoE factor. In order to showcase the generalization capability of our model, we conducted tests using different QoE weights. In addition, we tested the bitrate utilisation during video playback at different tile partition size settings. Finally, we conducted detailed ablation experiments to further investigate the individual contributions of various factors.
	
	\textbf{Viewport prediction:} In the viewport prediction section, we compare our method with the current state-of-the-art methods\cite{r36,r18,r11,r23} and the experimental results are shown in Fig. 5 and TABLE \uppercase\expandafter{\romannumeral 1}.
	\begin{figure}[!t]
		\centering
		\includegraphics[width=\columnwidth]{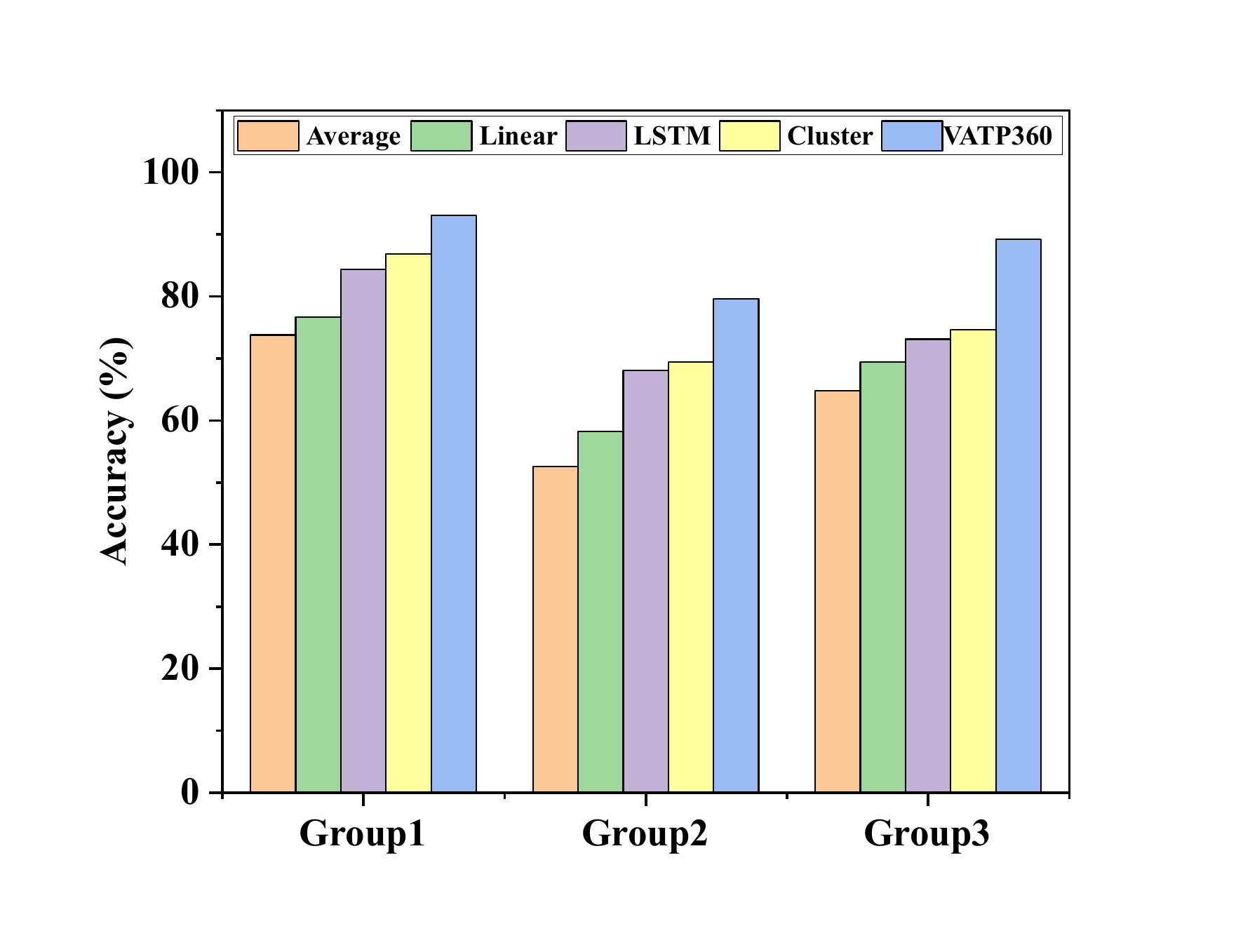}
		\captionsetup{format=plain, justification=justified}
		\caption{Comparison results of viewport prediction accuracy.}
		\label{fig_3}
	\end{figure}
	In our experiments, we categorized the 360-degree videos into three groups based on their content characteristics: Group1 consists of videos with static background, few objects in the foreground, and slow and single forms of motion. Group2 includes videos with more objects, as well as irregular and rapid forms of motion. Group3 comprises videos captured for outdoor activities, featuring objects such as pedestrians and buildings in daily life.

	\begin{table}[]
		\centering
		\caption{Accuracy of viewport prediction}
		\begin{tabular}{|c|c|c|c|}
			\hline
			Model & Group1 & Group2 & Group3 \\ \hline
			Average\cite{r36} & 73.8\% & 52.6\% & 64.8\% \\ \hline
			Linear\cite{r18} & 76.7\% & 58.2\% & 69.4\% \\ \hline
			LSTM\cite{r11} & 84.4\% & 68.1\% & 73.1\% \\ \hline
			Cluster\cite{r23} & 86.9\% & 69.4\% & 74.6\% \\ \hline
			\textbf{VATP360} & \textbf{93.1\%} & \textbf{79.6\%} & \textbf{89.2\%} \\ \hline
		\end{tabular}
	\end{table}
	
	The results show that our proposed viewport prediction method is on average 6\%-27\% more accurate than other baseline methods in terms of prediction accuracy. Because Average\cite{r35}, Linear\cite{r18} and LSTM\cite{r11} only use the user's historical viewport for prediction and do not take into account other factors that may affect changes in the user's viewport, it is ultimately difficult to achieve a high accuracy rate. Cluster\cite{r23} classifies and then clusters the viewports of users with different viewing preferences by prediction. However, this approach is non-adaptive and does not adapt well to changes in the user's viewport. Our method uses a joint prediction of the user's head movement trajectory and video saliency map, and adaptively adjusts the range of predicted viewports through secondary fitting by tile priority classification algorithm, finally achieving a prediction accuracy of 79.6\%-93.1\%.
	
	\begin{figure*}[!t]
		\centering
		\captionsetup[subfigure]{labelformat=empty}
		\subfloat[]{\includegraphics[width=2.5in]{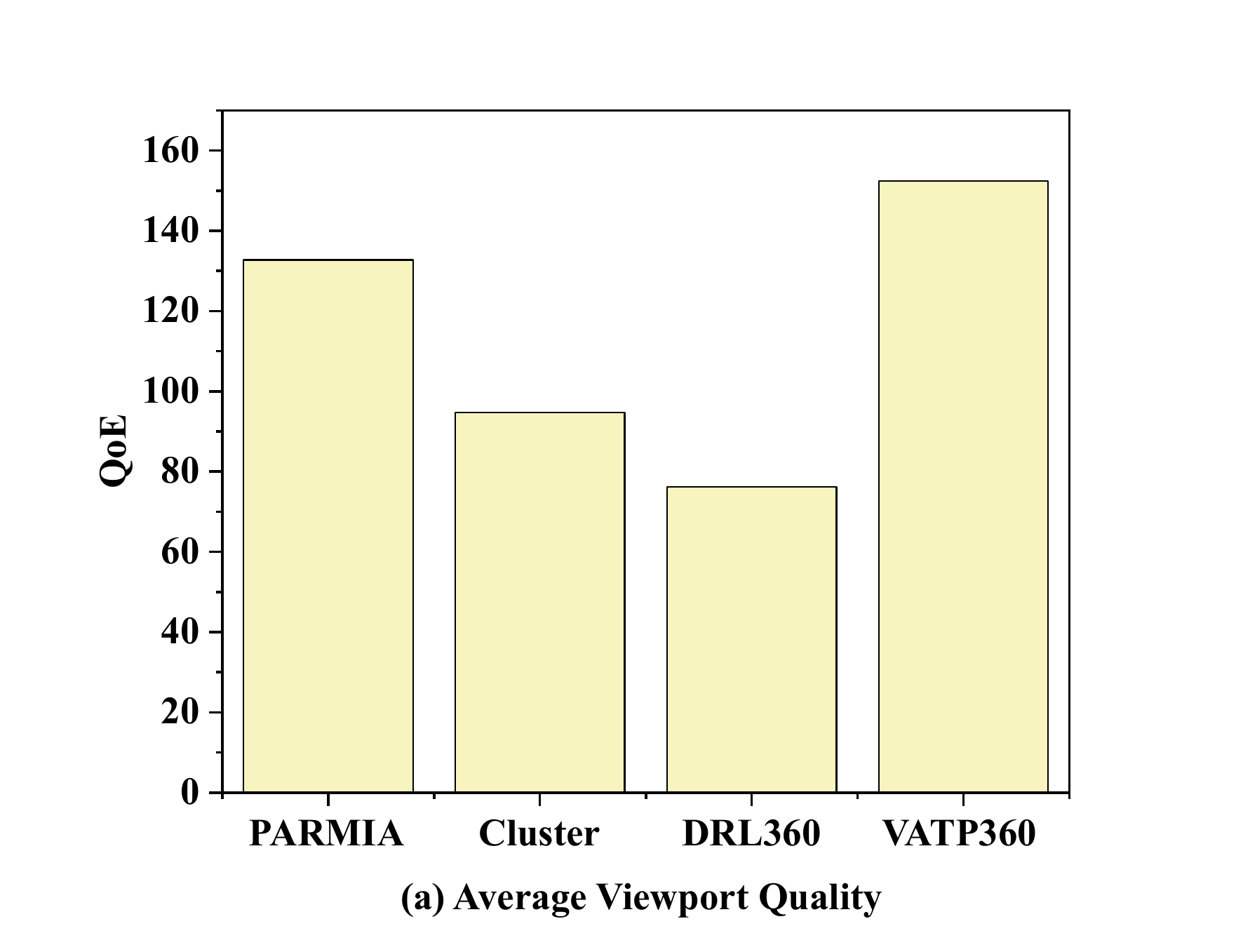}%
			\label{average viewport quality}}
		\hfil
		\subfloat[]{\includegraphics[width=2.5in]{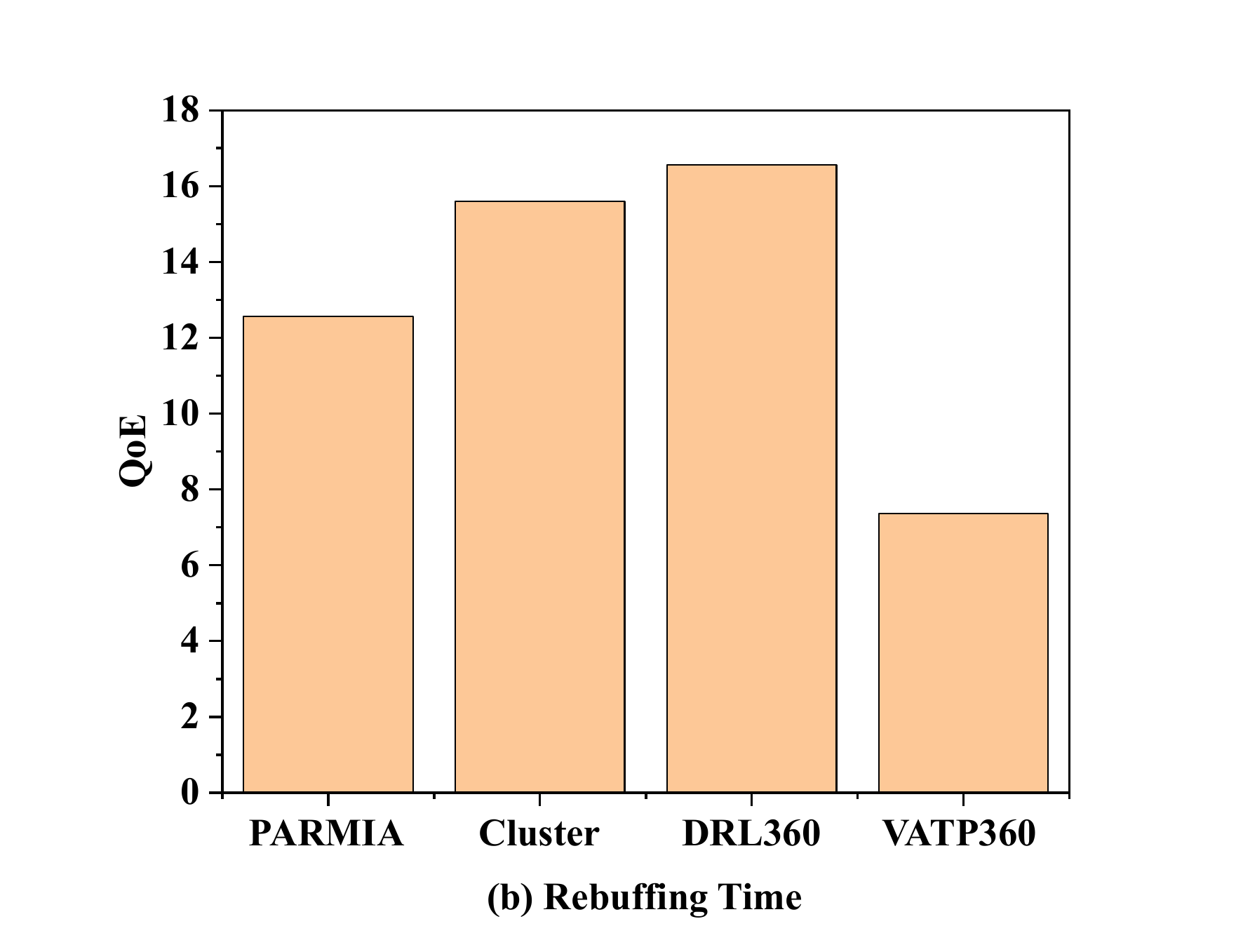}%
			\label{rebuffing time}}
		\hfil
		\subfloat[]{\includegraphics[width=2.5in]{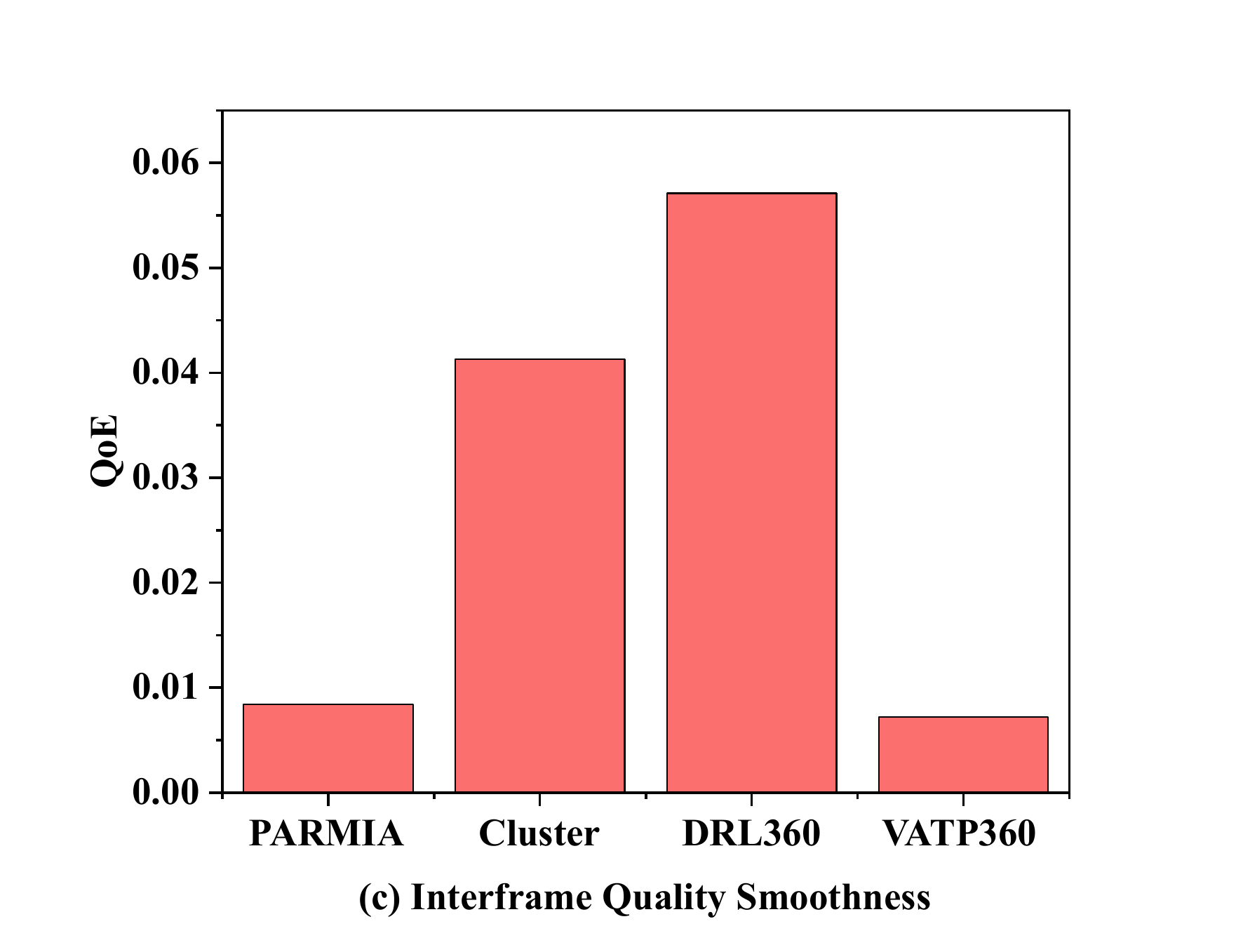}
			\label{interframe quality smoothness}}
		\hfil
		\subfloat[]{\includegraphics[width=2.5in]{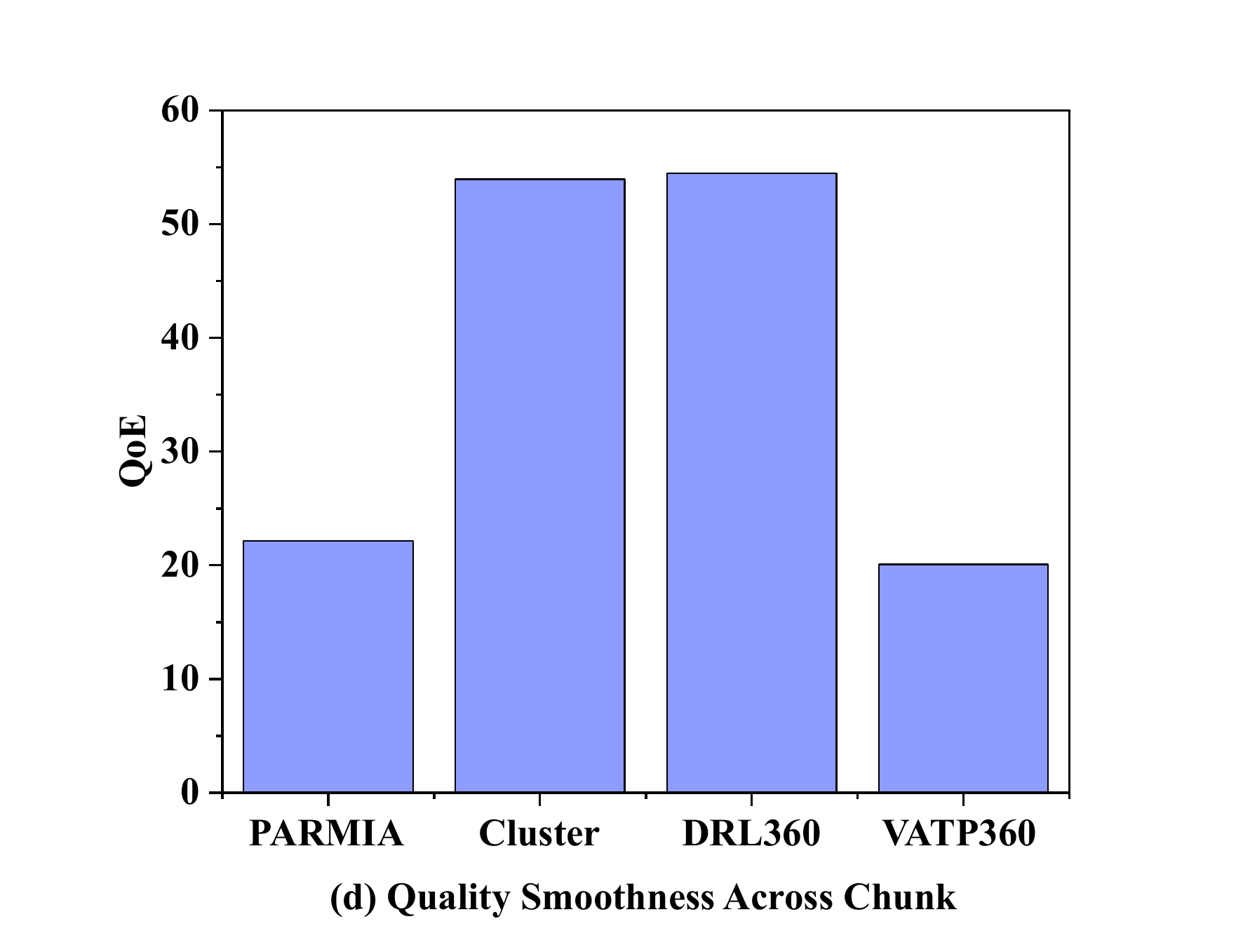}%
			\label{quality smoothness across chunk}}
		\caption{Performance comparison on each QoE component}
		\label{fig_4}
	\end{figure*}
	
	\textbf{Adaptive bitrate allocation:} For the bitrate allocation, we compare the effectiveness of our method with the current state-of-the-art methods for overall user QoE at different network bandwidths (4G network and the addition of 1Mbps and 2Mbps bandwidth) on two publicly available datasets\cite{r32,r33}.
	\begin{table}[]
		\centering
		\caption{Comparison of QoE under different bandwidth conditions}
		\begin{tabular}{|c|c|c|c|}
			\hline
			Model & 4G & +1Mbps & +2Mbps \\ \hline
			PARMIA\cite{r27} & 151.52 & 156.65 & 161.63 \\ \hline
			Cluster\cite{r23} & 111.11 & 115.16 & 118.58 \\ \hline
			Pensieve360 & 72.36 & 77.05 & 82.22 \\ \hline
			PanoSalNet\cite{r28} & 84.36 & 90.75 & 96.08 \\ \hline
			Flare\cite{r18} & 106.29 & 111.55 & 115.54 \\ \hline
			\textbf{VATP360} & \textbf{164.72} & \textbf{169.55} & \textbf{173.81} \\ \hline
		\end{tabular}
	\end{table}
	As can be seen from TABLE \uppercase\expandafter{\romannumeral 2}, our method outperforms other methods on both datasets, and our method has stable performance under different network bandwidth conditions. Among all, PARMIA\cite{r27} comes closest to our results as it uses the user's historical viewport and video object trajectory in viewport prediction, which can effectively predict the viewport. However, PARMIA assumes that the network bandwidth is constant, so it cannot adapt to real network bandwidth environments. Neither Cluster\cite{r23} nor PanoSalNet\cite{r28} take into account that the user viewport may be affected by the video content. Flare predicts the user viewport in a simple linear way. Pensieve360 follows the traditional video streaming approach for 360 degree video streams. As an additional note, in the later experiments, we set the network bandwidth condition to a 4G environment.
	
	We then tested the gain on each QoE component under the weights of (1, 1, 4, 4). The comparative results are shown in Fig. 6. The results show that our method outperforms the other methods on all four QoE factors. Although the performance of PARMIA\cite{r27} in terms of smoothness is close to our method, it is unable to adaptively cope with changes in the playback buffer and ends up experiencing longer re-buffering because it is trained under constant network bandwidth conditions. The other two adaptive schemes\cite{r11,r23} only consider a single effect of the user's historical viewport in the viewport prediction section and cannot adapt to changes in the user's viewport. It is difficult to achieve a high user QoE in terms of viewport quality and smoothness.
	
	\textbf{Tile partitioning:} We also tested the performance of our approach in terms of bitrate utilisation and user QoE under different tile partition approaches. For a single video frame, we calculate its bitrate utilisation by formula 12.
	\begin{equation}
		e=\frac{w_{top}b_{top}+w_{top-mid}b_{top-mid}}{\sum_{i=1}^{4}{w_{i}b_{i}}}
	\end{equation}
	where $b_{i}$ represents the bitrate assigned to the tile with priority $w_{i}$. We then calculate the bitrate utilisation for the entire 360 degree video playback period.
	\begin{equation}
		E=\frac{1}{mf_{m}}\sum_{i=1}^{m}\sum_{j=1}^{f_{m}}{e_{i,j}}
	\end{equation}
	
	The experimental results are shown in Fig. 7. 
	\begin{figure}[!t]
		\centering
		\includegraphics[width=\columnwidth]{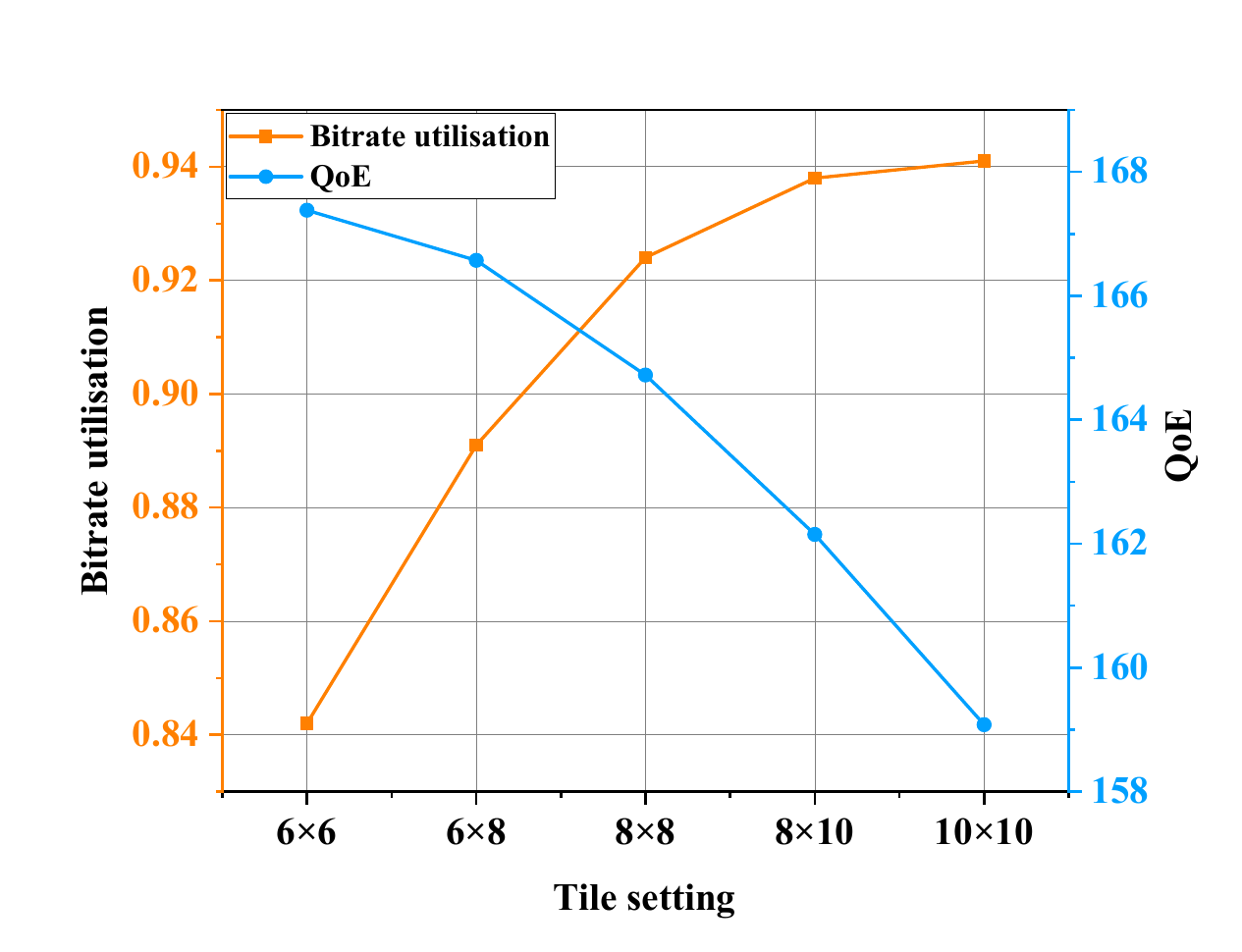}
		\captionsetup{format=plain, justification=justified} 
		\caption{Bitrate utilisation and user QoE performance for different tiling sizes}
		\label{fig_}
	\end{figure}
	When the whole video frame is tiled more finely, the number of tiles increases, and the model is able to allocate the bitrate more accurately for each video frame region, improving the bitrate utilisation during 360-degree video transmission. However, the higher the number of tiles, the more likely the viewport prediction is to be inaccurate, because when there is a slight change in the user's viewing behaviour, the viewport change may span across multiple tiles, leading to a degradation of the user's viewing experience. Moreover, the increase in tile will also lead to more redundancy in the video coding. So, we weighed the bitrate utilisation for 360-degree video transmission against the QoE for user viewing, and ultimately opted for an 8×8 tiled size.

	\textbf{QoE preferences:} In addition, we tested the results under different QoE weights. The three weight settings were considered as three groups of users with different viewing preferences. (1, 1, 1, 1) represents general users with no special viewing needs. (1, 4, 1, 1) represents users who focus on the smoothness of video playback, which requires fewer video lags during playback. (1, 1, 4, 4) represents users who focus on smoothness of video quality, with minimal fluctuations in video clarity during playback. As shown in Fig. 8, compared to other baseline methods, VATP360 can well meet the viewing needs of users with different preferences and has better generalisation ability.
	\begin{figure}[!t]
		\centering
		\includegraphics[width=\columnwidth]{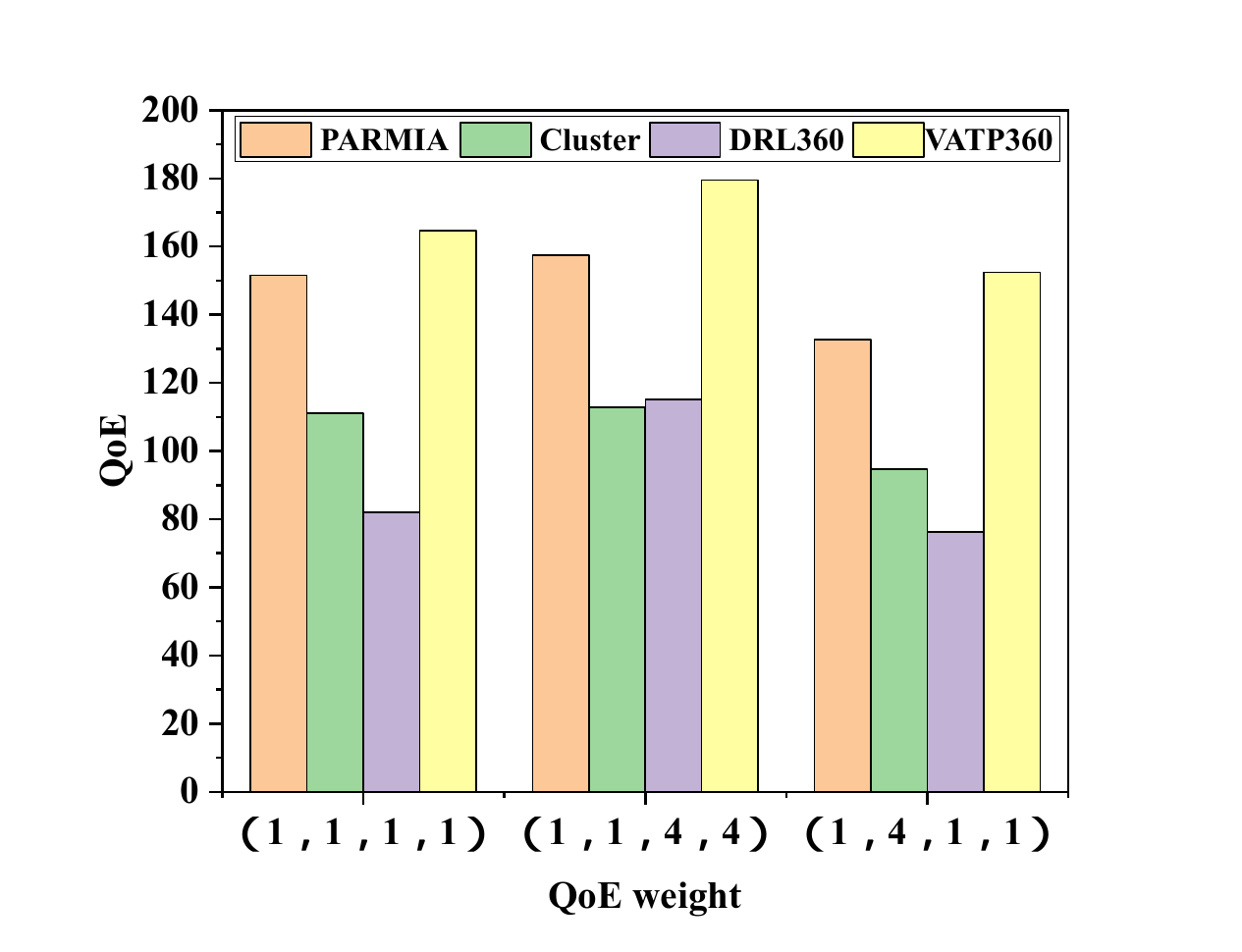}
		\captionsetup{format=plain, justification=justified} 
		\caption{Performance comparison under different QoE weights}
		\label{fig_}
	\end{figure}
	
	\textbf{Ablation study}: Finally, we conducted detailed ablation experiments on two publicly available datasets\cite{r32,r33} to demonstrate the effectiveness of the tile priority classification method proposed in this paper. In a controlled environment with 4G network bandwidth and weights set to (1, 1, 1, 1), we tested the performance of each module and calculated the normalized QoE using the model in this paper as a benchmark. The test results are shown in TABLE \uppercase\expandafter{\romannumeral 3}.
	
	\begin{table}[]
		\caption{Ablation study}
		\centering
		\begin{tabular}{|c|c|cl|}
			\hline
			Model & QoE & \multicolumn{2}{c|}{Normalized QoE} \\ \hline
			VP+PBA & 148.18 & \multicolumn{2}{c|}{89.88\%} \\ \hline
			VP+ABR & 157.16 & \multicolumn{2}{c|}{95.41\%} \\ \hline
			VP-s+ABR & 159.87 & \multicolumn{2}{c|}{97.06\%} \\ \hline
			\textbf{VATP360} & \textbf{164.72} & \multicolumn{2}{c|}{\textbf{100\%}} \\ \hline
		\end{tabular}
	\end{table}
	
	VP represents the viewport prediction module, using information from the video saliency map and the user's head movement trajectory as input to the LSTM network. VP-s represents our proposed viewport prediction module, which refine the VP prediction result with object detection. PBA is the non-adaptive bitrate allocation algorithm in\cite{r27}. ABR is the adaptive bitrate algorithm, which adaptively assigns the bitrate to each tile by reinforcement learning. Under unstable network bandwidth, it is difficult for PBA to provide a satisfactory viewing experience for users. The first two rows in TABLE \uppercase\expandafter{\romannumeral 3} demonstrate the efficiency of RL for 360 degree video streaming. Although ABR can adapt to fluctuations in network bandwidth, its performance can be severely affected when errors occur in the predicted viewport. The second and third rows in TABLE \uppercase\expandafter{\romannumeral 3}  demonstrate the effectiveness of the proposed viewport prediction module. Finally, the tile prioritisation algorithm in this paper not only effectively addresses errors in viewport prediction but also optimises the decision space of the RL adaptive algorithm, achieving a QoE gain of 2.94\%. The ablation study proves the effectiveness of our proposed viewport prediction algorithm and tile prioritisation algorithm.
	
	\section{conclusion}
	This paper investigates viewport prediction, tile priority classification, and bitrate allocation in 360-degree streaming, and proposes a novel adaptive 360-degree video streaming model VATP360. By combining object movement with user’s head movement trajectory and video saliency map, the accuracy of user's future viewport prediction is improved. The proposed tile priority-based classification algorithm further reduces viewport prediction errors and computational complexity of the model. Finally, bitrates are adaptively assigned to each tile using reinforcement learning model. Experimental results confirm the superiority of our proposed algorithm. 
	
	\section{Future work}
	In future work, we plan to consider more video-related features and different ways of tile partition, combining the content of the video with more detailed partition of the tiles to further improve the user QoE metrics.
	For the dataset, we consider using more user head movement data and 360-degree videos to enhance the generalisation capability of the model.

	\section*{Acknowledgments}
	This should be a simple paragraph before the References to thank those individuals and institutions who have supported your work on this article.

	\bibliographystyle{IEEEtran}
	\bibliography{reference}
	
	\newpage

	\vfill
	
\end{document}